\documentclass[11pt,a4paper,twoside]{book}
\usepackage{mylayout}
\usepackage{ifpdf}
\ifpdf
\usepackage{thumbpdf}
\RequirePackage[colorlinks,hyperindex,plainpages=false]{hyperref}
\else
\RequirePackage[plainpages=true]{hyperref}
\usepackage{color}
\fi

\usepackage{lshort}
\usepackage{makeidx,shortvrb,latexsym}

\makeindex
\begin{document}
\selectlanguage{english}
\frontmatter

%%%%%%%%%%%%%%%%%%%%%%%%%%%%%%%%%%%%%%%%%%%%%%%%%%%%%%%%%%%%%%%%%
% Contents: The title page
% $Id: title.tex 449 2010-12-14 16:53:51Z oetiker $
%%%%%%%%%%%%%%%%%%%%%%%%%%%%%%%%%%%%%%%%%%%%%%%%%%%%%%%%%%%%%%%%%

\ifpdf
  \pdfbookmark{Title Page}{title}
\fi
\newlength{\centeroffset}
\setlength{\centeroffset}{-0.5\oddsidemargin}
\addtolength{\centeroffset}{0.5\evensidemargin}
%\addtolength{\textwidth}{-\centeroffset}
\thispagestyle{empty}
\vspace*{\stretch{1}}
\noindent\hspace*{\centeroffset}\makebox[0pt][l]{\begin{minipage}{\textwidth}
{\Huge\bfseries Christhin}

\noindent\rule[-1ex]{\textwidth}{5pt}\\[2.5ex]
\hfill
\begin{flushright}
{\Large Quantitative Analysis of Thin Layer Chromatography \\ Version~1.00 for Christhin 0.1.36 \\March 2012}
\end{flushright}
\end{minipage}}

\vspace{\stretch{2}}

\begin{center}
		\begin{tabular}{cp{5.5cm}}
	\includegraphics[width=0.4\textwidth]{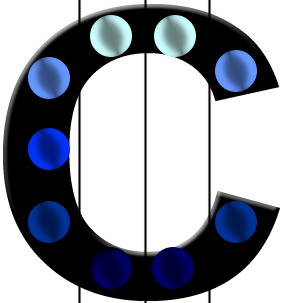} & {\Huge\bfseries Chromatography Riser Thin}
	\end{tabular}
\end{center}

\vspace{\stretch{5}}

\noindent\hspace*{\centeroffset}\makebox[0pt][l]{\begin{minipage}{\textwidth}
%\flushright
{\bfseries 
Maximiliano Barchiesi\\%[1.5ex]
Mercedes Sangroni\\%[1.5ex]
Carlos Renaudo\\%[1.5ex]
Pablo Rossi\\%[1.5ex]
Mar\'{i}a de Carmen Pramparo\\%[1.5ex]
Valeria Nepote\\%[1.5ex]
Nelson Ruben Grosso\\%[1.5ex]
Mar\'{i}a Fernanda Gayol\\[0ex]} 
\noindent\rule[-1ex]{\textwidth}{5pt}\\[2.5ex]
\hfill
\end{minipage}}

%\addtolength{\textwidth}{\centeroffset}
\vspace{\stretch{2}}

\pagebreak

\begin{small} 
  Copyright \copyright 2012 Barchiesi Maximiliano, Mercedes Sangroni, Carlos Renaudo, Pablo Rossi, Mar\'{i}a de Carmen Pramparo, Valeria Nepote, Nelson Ruben Grosso, Mar\'{i}a Fernanda Gayol.

This is the first edition of the Christhin documentation, and is consistent with version 0.1.36 of Christhin.

 Permission is granted to copy, distribute and/or modify this document under the terms of the GNU Free Documentation License, Version 1.3 or any later version published by the Free Software Foundation; with no Invariant Sections, no Front-Cover Texts, and no Back-Cover Texts.

   	A copy of the license is included in the section entitled ``GNU Free Documentation License''.

\end{small}

%%%%%%%%%%%%%%%%%%%%%%%%%%%%%%%%%%%%%%%%%%%%%%%%%%%%%%%%%%%%%%%%%
% Contents: Main Input File of the LaTeX2e Introduction
% $Id: lshort-base.tex 447 2010-12-14 14:32:00Z oetiker $
%%%%%%%%%%%%%%%%%%%%%%%%%%%%%%%%%%%%%%%%%%%%%%%%%%%%%%%%%%%%%%%%%
% lshort.tex - The not so short introduction to LaTeX   
%                                                      by Tobias Oetiker
%                                                     oetiker@ee.ethz.ch
%
%                           based on LKURTZ.TEX Uni Graz & TU Wien, 1987
%-----------------------------------------------------------------------
%
% To compile lshort, you need TeX 3.x, LaTeX and makeindex
%
% The sources files of the Intro are:
%      lshort.tex (this file),
%      titel.tex, contrib.tex, biblio.tex
%      things.tex, typeset.tex, math.tex, lssym.tex, spec.tex,
%      lshort.sty, fancyheadings.sty
%
% Further the  verbatim.sty and the layout.sty 
% from the LaTeX Tools distribution is
% required.
%
%
% To print the AMS symbols you need the AMS fonts and the packages
% amsfonts, eufrak and eucal from (AMS LaTeX 1.2)
%
% ---------------------------------------------------------------------

%%%%%%%%%%%%%%%%%%%%%%%%%%%%%%%%%%%%%%%%%%%%%%%%%%%%%%%%%%%%%%%%%
% Contents: Who contributed to this Document
% $Id: contrib.tex 449 2010-12-14 16:53:51Z oetiker $
%%%%%%%%%%%%%%%%%%%%%%%%%%%%%%%%%%%%%%%%%%%%%%%%%%%%%%%%%%%%%%%%%

\chapter{Authors}
\begin{verse}
\contrib{Maximiliano Barchiesi}{m_barchiesi@hotmail.com}%
{Universidad Nacional de R\'{i}o Cuarto}
\contrib{Mercedes Sangroni}{msangroni@hotmail.com.ar}%
   {Universidad Nacional de R\'{i}o Cuarto}
\contrib{Carlos Renaudo}{charlyrenaudo@hotmail.com}%
   {Universidad Nacional de R\'{i}o Cuarto}
\contrib{Pablo Rossi}{prossi@ing.unrc.edu.ar}%
   {Universidad Nacional de R\'{i}o Cuarto}
\end{verse}

\begin{verse}
\contrib{Mar\'{i}a de Carmen Pramparo}{mpramparo@ing.unrc.edu.ar}%
{Universidad Nacional de R\'{i}o Cuarto}
\contrib{Valeria Nepote}{vnepote@efn.uncor.edu}%
   {Universidad Nacional de C\'ordoba}
\contrib{Nelson Ruben Grosso}{nrgrosso@agro.uncor.edu}%
   {Universidad Nacional de C\'ordoba}
\contrib{Mar\'{i}a Fernanda Gayol}{mfgayol@ing.unrc.edu.ar}%
   {Universidad Nacional de R\'{i}o Cuarto}
\end{verse}

\chapter{Acknowledgments}

We thank CONICET, the Secretar\'{i}a de Ciencia, Tecnologia e Innovaci\'on, Gobierno de la Provincia de Chubut and the Consejo Interuniversitario Nacional (CIN) for supporting this development.

We also thank Dr. H\'ector ``Tito'' Fern\'andez and Dr. Nazareno Aguirre for their help.

% Local Variables:
% TeX-master: "lshort2e"
% mode: latex
% mode: flyspell
% End:

\include{overview}
\tableofcontents
%\listoffigures
%\listoftables
\enlargethispage{\baselineskip}
\mainmatter
%%%%%%%%%%%%%%%%%%%%%%%%%%%%%%%%%%%%%%%%%%%%%%%%%%%%%%%%%%%%%%%%%
% Contents: Things you need to know
% $Id: things.tex 448 2010-12-14 15:34:32Z oetiker $
%%%%%%%%%%%%%%%%%%%%%%%%%%%%%%%%%%%%%%%%%%%%%%%%%%%%%%%%%%%%%%%%%
 
\chapter{A brief introduction about Thin Layer Chromatography} 

Thin Layer Chromatography (TLC) is a chromatographic method on the plane. It employs a flat, thin layer, at the same time the support, and covers a surface (glass, plastic). Mobile phase or eluent  moves through the support or stationary phase by capillary action, allowing components of a mixture be separated by a difference in polarity. 

Is a fast, good resolution technique and with greater sensitivity than paper chromatography (allows detection of smaller amounts of analyte). This technique is widely used in industrial laboratories, and as a result of the many different application areas, it is estimated that as many TLC analysis are performed as HPLC analysis.

The advantages of following this procedure are speed and low cost of thin-layer experimental trials. The TLC is a useful tool for quality control and purity of products.

With respect to the final result of applying the TLC technique, comparing the relative intensities of the different spots presents, belonging each to a different compound, a quantitative estimate of the composition of the sample can be made. \cite{skoog01}

\section{Materials and methods}

Characteristics thin layer separations are performed on glass or plastic plates which are covered with a thin, adherent layer of finely divided particles; this layer forms the stationary phase. \cite{skoog01,stahl69}

\section{Plate Preparation}

Is generally used as stationary phase powdered silica gel, which is mixed with water in suitable proportions in order to prepare a homogeneous paste. It is distributed evenly over a glass plate and subjected to a drying oven for the time necessary, to remove humidity. The square glass plates to be used have a size of between $16$ to $20$ $cm$. sideways and seek a suitable stationary layer thickness, from $0.25$ to $0.75$ $mm$. approximately. \cite{aocs94}

\section{Preparation of the tray}

The tray must have appropriate measures in order to introduce inside it the plate where the run will be done; it is placed also inside it the solvent mixture serving as mobile phase. \cite{gunstone07}

As to the solvents used, alternatives may be varied, but generally are mixtures of hexane, di-ethyl-ether and acetic acid.

\section{Development}

The sample application is the most critical aspect of thin-layer chromatography, especially when dealing with quantitative measures. In general, the sample should be diluted to achieve a good advance and must be applied at $1$ or $2$ $cm$. from the lower end of the plate, thus avoiding to take direct contact with the solvent. For efficient separation, the patch must have a minimum diameter ($0.5$ $mm$ or less), so that the application is conducted trougth a capillary, or a specific chromatographic syringe.

During the test, the sample is transported by the mobile phase through the stationary phase by diffusion. For the assay, is marked with a pencil a line of $1$ or $2$ $cm$. from one end, and there is deposited the drop of the sample. Once the solvent to evaporate from the sample, we proceed to placing the plate in a closed vessel saturated with solvent vapors, which makes the assay (tray).

The end where the drop was deposited is introduced into the solvent avoiding direct contact; the mobile phase ascends the plate by capillary effect exerted between the fine particles of silica gel. When scrolling, the solvent passes through the point of application of sample, dissolves and drags it trough plate, distributing the sample among the mobile phase that is displaced and stationary phase according to the degree of polarity of the components. When the solvent reaches two-thirds of the length of the plate, this is removed from the tray and dried.

\section{Location of compounds of interest}

To locate the components on the board is sprayed over it a solution of iodine as it reacts with organic compounds to give dark products, or simply the board is placed in a tray similar to that containing the mixture of solvents, but containing instead solid iodine.

\section{Quantification}

For quantitative analysis of TLC chromatography, plate must be digitally processed by a digital-optical media, such as a scanner, and then process the information with an appropriate software. The software \textbf{Christhin} allows processing of images obtained by TLC, collecting information on the color intensity of the image, and relating this to the concentration of the compounds present in the sample analyzed.

\chapter{How to use Christhin}

To use the software Christhin, we recommend following the steps listed below:

\begin{enumerate}
	\item Install the program. This can be done in two different ways:

\begin{itemize}
	\item After installing the software Octave, from the main window of it run the command \emph{pkg install program name}, where instead of program name places the file name containing the Christhin's software, which is by default (the file downloaded from intenet) \emph{christhin-0.1.36.tar.gz}.
	
\item	After installing the software Octave, run the \emph{setup.exe} file that contains the material downloaded from internet, then follow the steps in the installation wizard, taking into account the detail of first specify the folder where you want to install the software, and then specify the folder where the software Octave is installed. Thus, the execution of the program allows immediate use of the software. 
\end{itemize}

\item	It is essential for the proper functioning of software, that the plate where chromatography was performed (once already revealed), is scanned and taken to image format type. This must be done prior to program execution.

\item	Run the software, which immediately asks you to select the image to be processed. After selecting the image, click \emph{Ok} to continue.

\item	If the scan of the plate was defective, the program allows rotating the image in order to make corrections; instead, not being necessary this correction is placed a rotation angle of zero degrees. The rotation must always be done by ensuring that the seed points of the sample remaining in the bottom of the image, position essential for the proper functioning of the software in the following steps.

\item	Select the rectangular area where the desired run is placed. This is done through a single click in the top left, and another on the opposite side, being formed an \textit{imaginary} rectangle. Remember to include in it both the solvent front as the point of sowing.

\item	After selecting the appropriate area, you press \emph{enter} to confirm. On screen will then appear a new image containing only the selected runing inside the \textit{imaginary} rectangle. On this image, select the seed point and the solvent front with a single click in each case. This step is intended to collect information to calculate the value \emph{Rf} (Ratio of Front). After selecting both two points (which can be selected interchangeably), press \emph{enter} to confirm.

\item	Screen now shows a chromatogram which contains the information of the run, being the peak area proportional to the intensity of the spots on the run, and therefore proportional to the amount of each component in the sample chromatographed. In it, select by clicking the start of each \emph{peak} and the end with another click, depending on how many of them possessing the image being analyzed. Make the selection without considering the height of the respective click (not give importance to the selected value on the y-axis, but the set value on the x-axis), as the program locates the click automatically over the curve.

\item	After selecting all the peaks, press \emph{enter}, and this completes the analysis of the run; the program lets you enter comments if necessary (if not carrying comments, press \emph{Ok} only).

\item	The program offers the possibility to analyze another run, because in general this kind of chromatography are performed several runs simultaneously on the same plate. If it is necessary to consider another run, select \emph{Yes}, and repeat steps 3 to 6 inclusive; not being necessary to consider another run, select \emph{No}, which ends the program execution.

\item As a result, in the same location of the processed image, the program creates a folder with the same name of the image, in which is saved the image in grayscale, the chromatogram which allowed the analysis, and a text file containing the comments entered, the percentage of area that represents each peak selected and the \emph{Rf} value for each compound. Each peak in the chromatogram has an assigned number, which corresponds to the number that contains the percentage of area in the text file.
 
\end{enumerate}

\chapter{Operation of the Software Christhin}
The software operation is carried out by a basic function, called \textbf{christhin} in collaboration with two other functions that have specific tasks, called \textbf{imgchop} and \textbf{pickdef}.

\section{Christhin}

This function is the main structure of the software, and the same is executed immediately upon being opened the program. The software asks the user to select the image to be processed, an image that contains the interest chromatography. The same is converted of a scale RGB, the color scale, to a grayscale image, thus facilitating further processing. Is allowed rotation of the image if needed, to ensure that seed points remaining always at the bottom of the image. The grayscale image is stored within a folder created by the program, with the same name that has the image analyzed.

Then the function \textbf{christhin} uses the function \textbf{imgchop} in order to obtain the area of interest where the run is going to be analyzed (see Section~\ref{pickdef}). Once generated the area of analysis, we proceed to select the seed point and the solvent front point, in a separate image created by the program for this purpose. In addition, the function generates based on the selected area the corresponding chromatogram. This is done taking into account the color intensity of spots in the image, and therefore uses the grayscale image.

Thus, the function \textbf{christhin} uses the function \textbf{pickdef} to calculate the percentage of each compound present in the sample, as well as their corresponding \emph{Rf} value (see Section~\ref{imgchop}).

Finally, through a number of commands, the program gives the images a specific format, color, font size and scale. Also save these images along with a text file in the folder mentioned above. The text file contains the percentage of each peak selected, and the \emph{Rf} value presented by each compound in the sample.

\section{Imgchop \label{imgchop}}

When used, the function \textbf{imgchop} allows the selection of the rectangular area in which is located the run that is going to be analyzed. The selection is done by clicking the top left and the other in the bottom right, forming an \textit{imaginary} rectangle, containing the running of interest, besides the seed point and solvent front. If the selection fails, the program issues an error message and asks again the selection. Thus, collects information about the location of the \textit{imaginary} rectangle where there is the run to be analyzed.

\section{Pickdef \label{pickdef}}

When used, the function sets first the output data, such as color and font size in graphics, as well as the nomenclature of each axis in the graphics created and their respective scales.

Then, selects the beginning and end of each peak in the chromatogram, previously generated within function \textbf{christhin}. The selection is made with a single click at the beginning and one at the end, and regardless of the y-axis value of the selection, but the x-axis value, because the program automatically places the selection on the curve of the chromatogram.
If the selection of the peaks is wrong, sometimes the program issues an error message, requesting the peaks to be selected again.

After selecting the peaks, the function calculates the area under the curve of each peak (using numerical integration known as the \textbf{trapezoid method}).

We also calculates the value of \emph{Rf} of each compound in the sample, making use of the values obtained when the user is prompted to mark the point of seeding and the front of solvent in the function \textbf{christhin}. (Remember that the value of \emph{Rf} is the ratio between the distance traveled by the compound of interest and the distance traveled by the solvent front, giving an idea of the degree of retention of the compound, directly related to its polarity).

Finally, it calculates the percentage of each compound in the analyzed sample, taking into account the percentage of area corresponding to each compound relative to the sum of all areas resulting from the selected peaks.

Thus, the function \textbf{pickdef} allows the calculation of the percentage of each compound in the sample, as well as the corresponding \emph{Rf} value.

\chapter{About Octave }

Text extracted from GNU Octave documentation. \cite{eaton08}

Octave is a high-level interactive language, primarily intended for numerical computations that is mostly compatible with Matlab\footnote{Matlab is a registered trademark of The MathWorks, Inc.}. Octave can do arithmetic for real, complex or integer-valued scalars and matrices, solve sets of nonlinear algebraic equations, integrate functions over finite and infinite intervals, and integrate systems of ordinary differential and differential-algebraic equations.

Octave uses the GNU readline library to handle reading and editing input. By default, the line editing commands are similar to the cursor movement commands used by GNU Emacs, and a vi-style line editing interface is also available. At the end of each session, the command history is saved, so that commands entered during previous sessions are not lost. The Octave distribution includes a 650+ page Texinfo manual. Access to the complete text of the manual is available via the doc command at the Octave prompt.

\section{Running Octave}

On most systems, Octave is started with the shell command \emph{octave}. Octave displays an initial message and then a prompt indicating it is ready to accept input. You can begin typing Octave commands immediately afterward. 

If you get into trouble, you can usually interrupt Octave by typing Control-C (written C-c for short). C-c gets its name from the fact that you type it by holding down CTRL and then pressing c. Doing this will normally return you to Octave's prompt. To exit Octave, type quit, or exit at the Octave prompt.

On systems that support job control, you can suspend Octave by sending it a SIGTSTP signal, usually by typing C-z.

\section{The platforms Octave}

Octave runs on various Unices-at least Linux and Solaris, Mac OS X, Windows and anything you can compile it on. Binary distributions exist at least for Debian, Suse, Fedora and RedHat Linuxes (Intel and AMD CPUs, at least), for Mac OS X and Windows' 98, 2000, XP, Vista, and 7.

Two and three dimensional plotting is fully supported using gnuplot and an experimental OpenGL backend.

The underlying numerical solvers are currently standard Fortran ones like LAPACK, LINPACK, ODEPACK, the BLAS, etc., packaged in a library of C++ classes. If possible, the Fortran subroutines are compiled with the system's Fortran compiler, and called directly from the C++ functions. If that's not possible, you can still compile Octave if you have the free Fortran to C translator f2c.

Octave is also free software; you can redistribute it and/or modify it under the terms of the GNU General Public License, version 3, as published by the Free Software Foundation, or at your option any later version.

\appendix
\appendix
\chapter{GNU GENERAL PUBLIC LICENSE}

\date{Version 3, 29 June 2007}

%\begin{document}
%\maketitle

\begin{center}
{\parindent 0in

Copyright \copyright\  2007 Free Software Foundation, Inc. \texttt{http://fsf.org/}

\bigskip
Everyone is permitted to copy and distribute verbatim copies of this

license document, but changing it is not allowed.}

\end{center}

%\renewcommand{\abstractname}{Preamble}

%\begin{abstract}

\begin{center}
{\Large \sc Preamble}
\end{center}

The GNU General Public License is a free, copyleft license for
software and other kinds of works.

The licenses for most software and other practical works are designed
to take away your freedom to share and change the works.  By contrast,
the GNU General Public License is intended to guarantee your freedom to
share and change all versions of a program--to make sure it remains free
software for all its users.  We, the Free Software Foundation, use the
GNU General Public License for most of our software; it applies also to
any other work released this way by its authors.  You can apply it to
your programs, too.

When we speak of free software, we are referring to freedom, not
price.  Our General Public Licenses are designed to make sure that you
have the freedom to distribute copies of free software (and charge for
them if you wish), that you receive source code or can get it if you
want it, that you can change the software or use pieces of it in new
free programs, and that you know you can do these things.

To protect your rights, we need to prevent others from denying you
these rights or asking you to surrender the rights.  Therefore, you have
certain responsibilities if you distribute copies of the software, or if
you modify it: responsibilities to respect the freedom of others.

For example, if you distribute copies of such a program, whether
gratis or for a fee, you must pass on to the recipients the same
freedoms that you received.  You must make sure that they, too, receive
or can get the source code.  And you must show them these terms so they
know their rights.

Developers that use the GNU GPL protect your rights with two steps:
(1) assert copyright on the software, and (2) offer you this License
giving you legal permission to copy, distribute and/or modify it.

For the developers' and authors' protection, the GPL clearly explains
that there is no warranty for this free software.  For both users' and
authors' sake, the GPL requires that modified versions be marked as
changed, so that their problems will not be attributed erroneously to
authors of previous versions.

Some devices are designed to deny users access to install or run
modified versions of the software inside them, although the manufacturer
can do so.  This is fundamentally incompatible with the aim of
protecting users' freedom to change the software.  The systematic
pattern of such abuse occurs in the area of products for individuals to
use, which is precisely where it is most unacceptable.  Therefore, we
have designed this version of the GPL to prohibit the practice for those
products.  If such problems arise substantially in other domains, we
stand ready to extend this provision to those domains in future versions
of the GPL, as needed to protect the freedom of users.

Finally, every program is threatened constantly by software patents.
States should not allow patents to restrict development and use of
software on general-purpose computers, but in those that do, we wish to
avoid the special danger that patents applied to a free program could
make it effectively proprietary.  To prevent this, the GPL assures that
patents cannot be used to render the program non-free.

The precise terms and conditions for copying, distribution and
modification follow.
%\end{abstract}

\begin{center}
{\Large \sc Terms and Conditions}
\end{center}

\begin{enumerate}

\addtocounter{enumi}{-1}

\item Definitions.

``This License'' refers to version 3 of the GNU General Public License.

``Copyright'' also means copyright-like laws that apply to other kinds of
works, such as semiconductor masks.

``The Program'' refers to any copyrightable work licensed under this
License.  Each licensee is addressed as ``you''.  ``Licensees'' and
``recipients'' may be individuals or organizations.

To ``modify'' a work means to copy from or adapt all or part of the work
in a fashion requiring copyright permission, other than the making of an
exact copy.  The resulting work is called a ``modified version'' of the
earlier work or a work ``based on'' the earlier work.

A ``covered work'' means either the unmodified Program or a work based
on the Program.

To ``propagate'' a work means to do anything with it that, without
permission, would make you directly or secondarily liable for
infringement under applicable copyright law, except executing it on a
computer or modifying a private copy.  Propagation includes copying,
distribution (with or without modification), making available to the
public, and in some countries other activities as well.

To ``convey'' a work means any kind of propagation that enables other
parties to make or receive copies.  Mere interaction with a user through
a computer network, with no transfer of a copy, is not conveying.

An interactive user interface displays ``Appropriate Legal Notices''
to the extent that it includes a convenient and prominently visible
feature that (1) displays an appropriate copyright notice, and (2)
tells the user that there is no warranty for the work (except to the
extent that warranties are provided), that licensees may convey the
work under this License, and how to view a copy of this License.  If
the interface presents a list of user commands or options, such as a
menu, a prominent item in the list meets this criterion.

\item Source Code.

The ``source code'' for a work means the preferred form of the work
for making modifications to it.  ``Object code'' means any non-source
form of a work.

A ``Standard Interface'' means an interface that either is an official
standard defined by a recognized standards body, or, in the case of
interfaces specified for a particular programming language, one that
is widely used among developers working in that language.

The ``System Libraries'' of an executable work include anything, other
than the work as a whole, that (a) is included in the normal form of
packaging a Major Component, but which is not part of that Major
Component, and (b) serves only to enable use of the work with that
Major Component, or to implement a Standard Interface for which an
implementation is available to the public in source code form.  A
``Major Component'', in this context, means a major essential component
(kernel, window system, and so on) of the specific operating system
(if any) on which the executable work runs, or a compiler used to
produce the work, or an object code interpreter used to run it.

The ``Corresponding Source'' for a work in object code form means all
the source code needed to generate, install, and (for an executable
work) run the object code and to modify the work, including scripts to
control those activities.  However, it does not include the work's
System Libraries, or general-purpose tools or generally available free
programs which are used unmodified in performing those activities but
which are not part of the work.  For example, Corresponding Source
includes interface definition files associated with source files for
the work, and the source code for shared libraries and dynamically
linked subprograms that the work is specifically designed to require,
such as by intimate data communication or control flow between those
subprograms and other parts of the work.

The Corresponding Source need not include anything that users
can regenerate automatically from other parts of the Corresponding
Source.

The Corresponding Source for a work in source code form is that
same work.

\item Basic Permissions.

All rights granted under this License are granted for the term of
copyright on the Program, and are irrevocable provided the stated
conditions are met.  This License explicitly affirms your unlimited
permission to run the unmodified Program.  The output from running a
covered work is covered by this License only if the output, given its
content, constitutes a covered work.  This License acknowledges your
rights of fair use or other equivalent, as provided by copyright law.

You may make, run and propagate covered works that you do not
convey, without conditions so long as your license otherwise remains
in force.  You may convey covered works to others for the sole purpose
of having them make modifications exclusively for you, or provide you
with facilities for running those works, provided that you comply with
the terms of this License in conveying all material for which you do
not control copyright.  Those thus making or running the covered works
for you must do so exclusively on your behalf, under your direction
and control, on terms that prohibit them from making any copies of
your copyrighted material outside their relationship with you.

Conveying under any other circumstances is permitted solely under
the conditions stated below.  Sublicensing is not allowed; section 10
makes it unnecessary.

\item Protecting Users' Legal Rights From Anti-Circumvention Law.

No covered work shall be deemed part of an effective technological
measure under any applicable law fulfilling obligations under article
11 of the WIPO copyright treaty adopted on 20 December 1996, or
similar laws prohibiting or restricting circumvention of such
measures.

When you convey a covered work, you waive any legal power to forbid
circumvention of technological measures to the extent such circumvention
is effected by exercising rights under this License with respect to
the covered work, and you disclaim any intention to limit operation or
modification of the work as a means of enforcing, against the work's
users, your or third parties' legal rights to forbid circumvention of
technological measures.

\item Conveying Verbatim Copies.

You may convey verbatim copies of the Program's source code as you
receive it, in any medium, provided that you conspicuously and
appropriately publish on each copy an appropriate copyright notice;
keep intact all notices stating that this License and any
non-permissive terms added in accord with section 7 apply to the code;
keep intact all notices of the absence of any warranty; and give all
recipients a copy of this License along with the Program.

You may charge any price or no price for each copy that you convey,
and you may offer support or warranty protection for a fee.

\item Conveying Modified Source Versions.

You may convey a work based on the Program, or the modifications to
produce it from the Program, in the form of source code under the
terms of section 4, provided that you also meet all of these conditions:
  \begin{enumerate}
  \item The work must carry prominent notices stating that you modified
  it, and giving a relevant date.

  \item The work must carry prominent notices stating that it is
  released under this License and any conditions added under section
  7.  This requirement modifies the requirement in section 4 to
  ``keep intact all notices''.

  \item You must license the entire work, as a whole, under this
  License to anyone who comes into possession of a copy.  This
  License will therefore apply, along with any applicable section 7
  additional terms, to the whole of the work, and all its parts,
  regardless of how they are packaged.  This License gives no
  permission to license the work in any other way, but it does not
  invalidate such permission if you have separately received it.

  \item If the work has interactive user interfaces, each must display
  Appropriate Legal Notices; however, if the Program has interactive
  interfaces that do not display Appropriate Legal Notices, your
  work need not make them do so.
\end{enumerate}
A compilation of a covered work with other separate and independent
works, which are not by their nature extensions of the covered work,
and which are not combined with it such as to form a larger program,
in or on a volume of a storage or distribution medium, is called an
``aggregate'' if the compilation and its resulting copyright are not
used to limit the access or legal rights of the compilation's users
beyond what the individual works permit.  Inclusion of a covered work
in an aggregate does not cause this License to apply to the other
parts of the aggregate.

\item Conveying Non-Source Forms.

You may convey a covered work in object code form under the terms
of sections 4 and 5, provided that you also convey the
machine-readable Corresponding Source under the terms of this License,
in one of these ways:
  \begin{enumerate}
  \item Convey the object code in, or embodied in, a physical product
  (including a physical distribution medium), accompanied by the
  Corresponding Source fixed on a durable physical medium
  customarily used for software interchange.

  \item Convey the object code in, or embodied in, a physical product
  (including a physical distribution medium), accompanied by a
  written offer, valid for at least three years and valid for as
  long as you offer spare parts or customer support for that product
  model, to give anyone who possesses the object code either (1) a
  copy of the Corresponding Source for all the software in the
  product that is covered by this License, on a durable physical
  medium customarily used for software interchange, for a price no
  more than your reasonable cost of physically performing this
  conveying of source, or (2) access to copy the
  Corresponding Source from a network server at no charge.

  \item Convey individual copies of the object code with a copy of the
  written offer to provide the Corresponding Source.  This
  alternative is allowed only occasionally and noncommercially, and
  only if you received the object code with such an offer, in accord
  with subsection 6b.

  \item Convey the object code by offering access from a designated
  place (gratis or for a charge), and offer equivalent access to the
  Corresponding Source in the same way through the same place at no
  further charge.  You need not require recipients to copy the
  Corresponding Source along with the object code.  If the place to
  copy the object code is a network server, the Corresponding Source
  may be on a different server (operated by you or a third party)
  that supports equivalent copying facilities, provided you maintain
  clear directions next to the object code saying where to find the
  Corresponding Source.  Regardless of what server hosts the
  Corresponding Source, you remain obligated to ensure that it is
  available for as long as needed to satisfy these requirements.

  \item Convey the object code using peer-to-peer transmission, provided
  you inform other peers where the object code and Corresponding
  Source of the work are being offered to the general public at no
  charge under subsection 6d.
  \end{enumerate}

A separable portion of the object code, whose source code is excluded
from the Corresponding Source as a System Library, need not be
included in conveying the object code work.

A ``User Product'' is either (1) a ``consumer product'', which means any
tangible personal property which is normally used for personal, family,
or household purposes, or (2) anything designed or sold for incorporation
into a dwelling.  In determining whether a product is a consumer product,
doubtful cases shall be resolved in favor of coverage.  For a particular
product received by a particular user, ``normally used'' refers to a
typical or common use of that class of product, regardless of the status
of the particular user or of the way in which the particular user
actually uses, or expects or is expected to use, the product.  A product
is a consumer product regardless of whether the product has substantial
commercial, industrial or non-consumer uses, unless such uses represent
the only significant mode of use of the product.

``Installation Information'' for a User Product means any methods,
procedures, authorization keys, or other information required to install
and execute modified versions of a covered work in that User Product from
a modified version of its Corresponding Source.  The information must
suffice to ensure that the continued functioning of the modified object
code is in no case prevented or interfered with solely because
modification has been made.

If you convey an object code work under this section in, or with, or
specifically for use in, a User Product, and the conveying occurs as
part of a transaction in which the right of possession and use of the
User Product is transferred to the recipient in perpetuity or for a
fixed term (regardless of how the transaction is characterized), the
Corresponding Source conveyed under this section must be accompanied
by the Installation Information.  But this requirement does not apply
if neither you nor any third party retains the ability to install
modified object code on the User Product (for example, the work has
been installed in ROM).

The requirement to provide Installation Information does not include a
requirement to continue to provide support service, warranty, or updates
for a work that has been modified or installed by the recipient, or for
the User Product in which it has been modified or installed.  Access to a
network may be denied when the modification itself materially and
adversely affects the operation of the network or violates the rules and
protocols for communication across the network.

Corresponding Source conveyed, and Installation Information provided,
in accord with this section must be in a format that is publicly
documented (and with an implementation available to the public in
source code form), and must require no special password or key for
unpacking, reading or copying.

\item Additional Terms.

``Additional permissions'' are terms that supplement the terms of this
License by making exceptions from one or more of its conditions.
Additional permissions that are applicable to the entire Program shall
be treated as though they were included in this License, to the extent
that they are valid under applicable law.  If additional permissions
apply only to part of the Program, that part may be used separately
under those permissions, but the entire Program remains governed by
this License without regard to the additional permissions.

When you convey a copy of a covered work, you may at your option
remove any additional permissions from that copy, or from any part of
it.  (Additional permissions may be written to require their own
removal in certain cases when you modify the work.)  You may place
additional permissions on material, added by you to a covered work,
for which you have or can give appropriate copyright permission.

Notwithstanding any other provision of this License, for material you
add to a covered work, you may (if authorized by the copyright holders of
that material) supplement the terms of this License with terms:
  \begin{enumerate}
  \item Disclaiming warranty or limiting liability differently from the
  terms of sections 15 and 16 of this License; or

  \item Requiring preservation of specified reasonable legal notices or
  author attributions in that material or in the Appropriate Legal
  Notices displayed by works containing it; or

  \item Prohibiting misrepresentation of the origin of that material, or
  requiring that modified versions of such material be marked in
  reasonable ways as different from the original version; or

  \item Limiting the use for publicity purposes of names of licensors or
  authors of the material; or

  \item Declining to grant rights under trademark law for use of some
  trade names, trademarks, or service marks; or

  \item Requiring indemnification of licensors and authors of that
  material by anyone who conveys the material (or modified versions of
  it) with contractual assumptions of liability to the recipient, for
  any liability that these contractual assumptions directly impose on
  those licensors and authors.
  \end{enumerate}

All other non-permissive additional terms are considered ``further
restrictions'' within the meaning of section 10.  If the Program as you
received it, or any part of it, contains a notice stating that it is
governed by this License along with a term that is a further
restriction, you may remove that term.  If a license document contains
a further restriction but permits relicensing or conveying under this
License, you may add to a covered work material governed by the terms
of that license document, provided that the further restriction does
not survive such relicensing or conveying.

If you add terms to a covered work in accord with this section, you
must place, in the relevant source files, a statement of the
additional terms that apply to those files, or a notice indicating
where to find the applicable terms.

Additional terms, permissive or non-permissive, may be stated in the
form of a separately written license, or stated as exceptions;
the above requirements apply either way.

\item Termination.

You may not propagate or modify a covered work except as expressly
provided under this License.  Any attempt otherwise to propagate or
modify it is void, and will automatically terminate your rights under
this License (including any patent licenses granted under the third
paragraph of section 11).

However, if you cease all violation of this License, then your
license from a particular copyright holder is reinstated (a)
provisionally, unless and until the copyright holder explicitly and
finally terminates your license, and (b) permanently, if the copyright
holder fails to notify you of the violation by some reasonable means
prior to 60 days after the cessation.

Moreover, your license from a particular copyright holder is
reinstated permanently if the copyright holder notifies you of the
violation by some reasonable means, this is the first time you have
received notice of violation of this License (for any work) from that
copyright holder, and you cure the violation prior to 30 days after
your receipt of the notice.

Termination of your rights under this section does not terminate the
licenses of parties who have received copies or rights from you under
this License.  If your rights have been terminated and not permanently
reinstated, you do not qualify to receive new licenses for the same
material under section 10.

\item Acceptance Not Required for Having Copies.

You are not required to accept this License in order to receive or
run a copy of the Program.  Ancillary propagation of a covered work
occurring solely as a consequence of using peer-to-peer transmission
to receive a copy likewise does not require acceptance.  However,
nothing other than this License grants you permission to propagate or
modify any covered work.  These actions infringe copyright if you do
not accept this License.  Therefore, by modifying or propagating a
covered work, you indicate your acceptance of this License to do so.

\item Automatic Licensing of Downstream Recipients.

Each time you convey a covered work, the recipient automatically
receives a license from the original licensors, to run, modify and
propagate that work, subject to this License.  You are not responsible
for enforcing compliance by third parties with this License.

An ``entity transaction'' is a transaction transferring control of an
organization, or substantially all assets of one, or subdividing an
organization, or merging organizations.  If propagation of a covered
work results from an entity transaction, each party to that
transaction who receives a copy of the work also receives whatever
licenses to the work the party's predecessor in interest had or could
give under the previous paragraph, plus a right to possession of the
Corresponding Source of the work from the predecessor in interest, if
the predecessor has it or can get it with reasonable efforts.

You may not impose any further restrictions on the exercise of the
rights granted or affirmed under this License.  For example, you may
not impose a license fee, royalty, or other charge for exercise of
rights granted under this License, and you may not initiate litigation
(including a cross-claim or counterclaim in a lawsuit) alleging that
any patent claim is infringed by making, using, selling, offering for
sale, or importing the Program or any portion of it.

\item Patents.

A ``contributor'' is a copyright holder who authorizes use under this
License of the Program or a work on which the Program is based.  The
work thus licensed is called the contributor's ``contributor version''.

A contributor's ``essential patent claims'' are all patent claims
owned or controlled by the contributor, whether already acquired or
hereafter acquired, that would be infringed by some manner, permitted
by this License, of making, using, or selling its contributor version,
but do not include claims that would be infringed only as a
consequence of further modification of the contributor version.  For
purposes of this definition, ``control'' includes the right to grant
patent sublicenses in a manner consistent with the requirements of
this License.

Each contributor grants you a non-exclusive, worldwide, royalty-free
patent license under the contributor's essential patent claims, to
make, use, sell, offer for sale, import and otherwise run, modify and
propagate the contents of its contributor version.

In the following three paragraphs, a ``patent license'' is any express
agreement or commitment, however denominated, not to enforce a patent
(such as an express permission to practice a patent or covenant not to
sue for patent infringement).  To ``grant'' such a patent license to a
party means to make such an agreement or commitment not to enforce a
patent against the party.

If you convey a covered work, knowingly relying on a patent license,
and the Corresponding Source of the work is not available for anyone
to copy, free of charge and under the terms of this License, through a
publicly available network server or other readily accessible means,
then you must either (1) cause the Corresponding Source to be so
available, or (2) arrange to deprive yourself of the benefit of the
patent license for this particular work, or (3) arrange, in a manner
consistent with the requirements of this License, to extend the patent
license to downstream recipients.  ``Knowingly relying'' means you have
actual knowledge that, but for the patent license, your conveying the
covered work in a country, or your recipient's use of the covered work
in a country, would infringe one or more identifiable patents in that
country that you have reason to believe are valid.

If, pursuant to or in connection with a single transaction or
arrangement, you convey, or propagate by procuring conveyance of, a
covered work, and grant a patent license to some of the parties
receiving the covered work authorizing them to use, propagate, modify
or convey a specific copy of the covered work, then the patent license
you grant is automatically extended to all recipients of the covered
work and works based on it.

A patent license is ``discriminatory'' if it does not include within
the scope of its coverage, prohibits the exercise of, or is
conditioned on the non-exercise of one or more of the rights that are
specifically granted under this License.  You may not convey a covered
work if you are a party to an arrangement with a third party that is
in the business of distributing software, under which you make payment
to the third party based on the extent of your activity of conveying
the work, and under which the third party grants, to any of the
parties who would receive the covered work from you, a discriminatory
patent license (a) in connection with copies of the covered work
conveyed by you (or copies made from those copies), or (b) primarily
for and in connection with specific products or compilations that
contain the covered work, unless you entered into that arrangement,
or that patent license was granted, prior to 28 March 2007.

Nothing in this License shall be construed as excluding or limiting
any implied license or other defenses to infringement that may
otherwise be available to you under applicable patent law.

\item No Surrender of Others' Freedom.

If conditions are imposed on you (whether by court order, agreement or
otherwise) that contradict the conditions of this License, they do not
excuse you from the conditions of this License.  If you cannot convey a
covered work so as to satisfy simultaneously your obligations under this
License and any other pertinent obligations, then as a consequence you may
not convey it at all.  For example, if you agree to terms that obligate you
to collect a royalty for further conveying from those to whom you convey
the Program, the only way you could satisfy both those terms and this
License would be to refrain entirely from conveying the Program.

\item Use with the GNU Affero General Public License.

Notwithstanding any other provision of this License, you have
permission to link or combine any covered work with a work licensed
under version 3 of the GNU Affero General Public License into a single
combined work, and to convey the resulting work.  The terms of this
License will continue to apply to the part which is the covered work,
but the special requirements of the GNU Affero General Public License,
section 13, concerning interaction through a network will apply to the
combination as such.

\item Revised Versions of this License.

The Free Software Foundation may publish revised and/or new versions of
the GNU General Public License from time to time.  Such new versions will
be similar in spirit to the present version, but may differ in detail to
address new problems or concerns.

Each version is given a distinguishing version number.  If the
Program specifies that a certain numbered version of the GNU General
Public License ``or any later version'' applies to it, you have the
option of following the terms and conditions either of that numbered
version or of any later version published by the Free Software
Foundation.  If the Program does not specify a version number of the
GNU General Public License, you may choose any version ever published
by the Free Software Foundation.

If the Program specifies that a proxy can decide which future
versions of the GNU General Public License can be used, that proxy's
public statement of acceptance of a version permanently authorizes you
to choose that version for the Program.

Later license versions may give you additional or different
permissions.  However, no additional obligations are imposed on any
author or copyright holder as a result of your choosing to follow a
later version.

\item Disclaimer of Warranty.

\begin{sloppypar}
 THERE IS NO WARRANTY FOR THE PROGRAM, TO THE EXTENT PERMITTED BY
 APPLICABLE LAW.  EXCEPT WHEN OTHERWISE STATED IN WRITING THE
 COPYRIGHT HOLDERS AND/OR OTHER PARTIES PROVIDE THE PROGRAM ``AS IS''
 WITHOUT WARRANTY OF ANY KIND, EITHER EXPRESSED OR IMPLIED,
 INCLUDING, BUT NOT LIMITED TO, THE IMPLIED WARRANTIES OF
 MERCHANTABILITY AND FITNESS FOR A PARTICULAR PURPOSE.  THE ENTIRE
 RISK AS TO THE QUALITY AND PERFORMANCE OF THE PROGRAM IS WITH YOU.
 SHOULD THE PROGRAM PROVE DEFECTIVE, YOU ASSUME THE COST OF ALL
 NECESSARY SERVICING, REPAIR OR CORRECTION.
\end{sloppypar}

\item Limitation of Liability.

 IN NO EVENT UNLESS REQUIRED BY APPLICABLE LAW OR AGREED TO IN
 WRITING WILL ANY COPYRIGHT HOLDER, OR ANY OTHER PARTY WHO MODIFIES
 AND/OR CONVEYS THE PROGRAM AS PERMITTED ABOVE, BE LIABLE TO YOU FOR
 DAMAGES, INCLUDING ANY GENERAL, SPECIAL, INCIDENTAL OR CONSEQUENTIAL
 DAMAGES ARISING OUT OF THE USE OR INABILITY TO USE THE PROGRAM
 (INCLUDING BUT NOT LIMITED TO LOSS OF DATA OR DATA BEING RENDERED
 INACCURATE OR LOSSES SUSTAINED BY YOU OR THIRD PARTIES OR A FAILURE
 OF THE PROGRAM TO OPERATE WITH ANY OTHER PROGRAMS), EVEN IF SUCH
 HOLDER OR OTHER PARTY HAS BEEN ADVISED OF THE POSSIBILITY OF SUCH
 DAMAGES.

\item Interpretation of Sections 15 and 16.

If the disclaimer of warranty and limitation of liability provided
above cannot be given local legal effect according to their terms,
reviewing courts shall apply local law that most closely approximates
an absolute waiver of all civil liability in connection with the
Program, unless a warranty or assumption of liability accompanies a
copy of the Program in return for a fee.

\begin{center}
{\Large\sc End of Terms and Conditions}
\end{center}

\end{enumerate}
%\end{document}

% This is set up to run with pdflatex.
%---------The file header---------------------------------------------

\hfuzz = .6pt % avoid black boxes
           
%---------------------------------------------------------------------
\chapter*{\rlap{GNU Free Documentation License}}
\phantomsection  % so hyperref creates bookmarks
\addcontentsline{toc}{chapter}{GNU Free Documentation License}
%\label{label_fdl}

 \begin{center}

       Version 1.3, 3 November 2008

 Copyright \copyright{} 2000, 2001, 2002, 2007, 2008  Free Software Foundation, Inc.
 
 \bigskip
 
     \texttt{<http://fsf.org/>}
  
 \bigskip
 
 Everyone is permitted to copy and distribute verbatim copies
 of this license document, but changing it is not allowed.
\end{center}

\begin{center}
{\bf\large Preamble}
\end{center}

The purpose of this License is to make a manual, textbook, or other
functional and useful document ``free'' in the sense of freedom: to
assure everyone the effective freedom to copy and redistribute it,
with or without modifying it, either commercially or noncommercially.
Secondarily, this License preserves for the author and publisher a way
to get credit for their work, while not being considered responsible
for modifications made by others.

This License is a kind of ``copyleft'', which means that derivative
works of the document must themselves be free in the same sense.  It
complements the GNU General Public License, which is a copyleft
license designed for free software.

We have designed this License in order to use it for manuals for free
software, because free software needs free documentation: a free
program should come with manuals providing the same freedoms that the
software does.  But this License is not limited to software manuals;
it can be used for any textual work, regardless of subject matter or
whether it is published as a printed book.  We recommend this License
principally for works whose purpose is instruction or reference.

\begin{center}
{\Large\bf 1. APPLICABILITY AND DEFINITIONS\par}
\phantomsection
\addcontentsline{toc}{section}{1. APPLICABILITY AND DEFINITIONS}
\end{center}

This License applies to any manual or other work, in any medium, that
contains a notice placed by the copyright holder saying it can be
distributed under the terms of this License.  Such a notice grants a
world-wide, royalty-free license, unlimited in duration, to use that
work under the conditions stated herein.  The ``\textbf{Document}'', below,
refers to any such manual or work.  Any member of the public is a
licensee, and is addressed as ``\textbf{you}''.  You accept the license if you
copy, modify or distribute the work in a way requiring permission
under copyright law.

A ``\textbf{Modified Version}'' of the Document means any work containing the
Document or a portion of it, either copied verbatim, or with
modifications and/or translated into another language.

A ``\textbf{Secondary Section}'' is a named appendix or a front-matter section of
the Document that deals exclusively with the relationship of the
publishers or authors of the Document to the Document's overall subject
(or to related matters) and contains nothing that could fall directly
within that overall subject.  (Thus, if the Document is in part a
textbook of mathematics, a Secondary Section may not explain any
mathematics.)  The relationship could be a matter of historical
connection with the subject or with related matters, or of legal,
commercial, philosophical, ethical or political position regarding
them.

The ``\textbf{Invariant Sections}'' are certain Secondary Sections whose titles
are designated, as being those of Invariant Sections, in the notice
that says that the Document is released under this License.  If a
section does not fit the above definition of Secondary then it is not
allowed to be designated as Invariant.  The Document may contain zero
Invariant Sections.  If the Document does not identify any Invariant
Sections then there are none.

The ``\textbf{Cover Texts}'' are certain short passages of text that are listed,
as Front-Cover Texts or Back-Cover Texts, in the notice that says that
the Document is released under this License.  A Front-Cover Text may
be at most 5 words, and a Back-Cover Text may be at most 25 words.

A ``\textbf{Transparent}'' copy of the Document means a machine-readable copy,
represented in a format whose specification is available to the
general public, that is suitable for revising the document
straightforwardly with generic text editors or (for images composed of
pixels) generic paint programs or (for drawings) some widely available
drawing editor, and that is suitable for input to text formatters or
for automatic translation to a variety of formats suitable for input
to text formatters.  A copy made in an otherwise Transparent file
format whose markup, or absence of markup, has been arranged to thwart
or discourage subsequent modification by readers is not Transparent.
An image format is not Transparent if used for any substantial amount
of text.  A copy that is not ``Transparent'' is called ``\textbf{Opaque}''.

Examples of suitable formats for Transparent copies include plain
ASCII without markup, Texinfo input format, LaTeX input format, SGML
or XML using a publicly available DTD, and standard-conforming simple
HTML, PostScript or PDF designed for human modification.  Examples of
transparent image formats include PNG, XCF and JPG.  Opaque formats
include proprietary formats that can be read and edited only by
proprietary word processors, SGML or XML for which the DTD and/or
processing tools are not generally available, and the
machine-generated HTML, PostScript or PDF produced by some word
processors for output purposes only.

The ``\textbf{Title Page}'' means, for a printed book, the title page itself,
plus such following pages as are needed to hold, legibly, the material
this License requires to appear in the title page.  For works in
formats which do not have any title page as such, ``Title Page'' means
the text near the most prominent appearance of the work's title,
preceding the beginning of the body of the text.

The ``\textbf{publisher}'' means any person or entity that distributes
copies of the Document to the public.

A section ``\textbf{Entitled XYZ}'' means a named subunit of the Document whose
title either is precisely XYZ or contains XYZ in parentheses following
text that translates XYZ in another language.  (Here XYZ stands for a
specific section name mentioned below, such as ``\textbf{Acknowledgements}'',
``\textbf{Dedications}'', ``\textbf{Endorsements}'', or ``\textbf{History}''.)  
To ``\textbf{Preserve the Title}''
of such a section when you modify the Document means that it remains a
section ``Entitled XYZ'' according to this definition.

The Document may include Warranty Disclaimers next to the notice which
states that this License applies to the Document.  These Warranty
Disclaimers are considered to be included by reference in this
License, but only as regards disclaiming warranties: any other
implication that these Warranty Disclaimers may have is void and has
no effect on the meaning of this License.

\begin{center}
{\Large\bf 2. VERBATIM COPYING\par}
\phantomsection
\addcontentsline{toc}{section}{2. VERBATIM COPYING}
\end{center}

You may copy and distribute the Document in any medium, either
commercially or noncommercially, provided that this License, the
copyright notices, and the license notice saying this License applies
to the Document are reproduced in all copies, and that you add no other
conditions whatsoever to those of this License.  You may not use
technical measures to obstruct or control the reading or further
copying of the copies you make or distribute.  However, you may accept
compensation in exchange for copies.  If you distribute a large enough
number of copies you must also follow the conditions in section~3.

You may also lend copies, under the same conditions stated above, and
you may publicly display copies.

\begin{center}
{\Large\bf 3. COPYING IN QUANTITY\par}
\phantomsection
\addcontentsline{toc}{section}{3. COPYING IN QUANTITY}
\end{center}

If you publish printed copies (or copies in media that commonly have
printed covers) of the Document, numbering more than 100, and the
Document's license notice requires Cover Texts, you must enclose the
copies in covers that carry, clearly and legibly, all these Cover
Texts: Front-Cover Texts on the front cover, and Back-Cover Texts on
the back cover.  Both covers must also clearly and legibly identify
you as the publisher of these copies.  The front cover must present
the full title with all words of the title equally prominent and
visible.  You may add other material on the covers in addition.
Copying with changes limited to the covers, as long as they preserve
the title of the Document and satisfy these conditions, can be treated
as verbatim copying in other respects.

If the required texts for either cover are too voluminous to fit
legibly, you should put the first ones listed (as many as fit
reasonably) on the actual cover, and continue the rest onto adjacent
pages.

If you publish or distribute Opaque copies of the Document numbering
more than 100, you must either include a machine-readable Transparent
copy along with each Opaque copy, or state in or with each Opaque copy
a computer-network location from which the general network-using
public has access to download using public-standard network protocols
a complete Transparent copy of the Document, free of added material.
If you use the latter option, you must take reasonably prudent steps,
when you begin distribution of Opaque copies in quantity, to ensure
that this Transparent copy will remain thus accessible at the stated
location until at least one year after the last time you distribute an
Opaque copy (directly or through your agents or retailers) of that
edition to the public.

It is requested, but not required, that you contact the authors of the
Document well before redistributing any large number of copies, to give
them a chance to provide you with an updated version of the Document.

\begin{center}
{\Large\bf 4. MODIFICATIONS\par}
\phantomsection
\addcontentsline{toc}{section}{4. MODIFICATIONS}
\end{center}

You may copy and distribute a Modified Version of the Document under
the conditions of sections 2 and 3 above, provided that you release
the Modified Version under precisely this License, with the Modified
Version filling the role of the Document, thus licensing distribution
and modification of the Modified Version to whoever possesses a copy
of it.  In addition, you must do these things in the Modified Version:

\begin{itemize}
\item[A.] 
   Use in the Title Page (and on the covers, if any) a title distinct
   from that of the Document, and from those of previous versions
   (which should, if there were any, be listed in the History section
   of the Document).  You may use the same title as a previous version
   if the original publisher of that version gives permission.
   
\item[B.]
   List on the Title Page, as authors, one or more persons or entities
   responsible for authorship of the modifications in the Modified
   Version, together with at least five of the principal authors of the
   Document (all of its principal authors, if it has fewer than five),
   unless they release you from this requirement.
   
\item[C.]
   State on the Title page the name of the publisher of the
   Modified Version, as the publisher.
   
\item[D.]
   Preserve all the copyright notices of the Document.
   
\item[E.]
   Add an appropriate copyright notice for your modifications
   adjacent to the other copyright notices.
   
\item[F.]
   Include, immediately after the copyright notices, a license notice
   giving the public permission to use the Modified Version under the
   terms of this License, in the form shown in the Addendum below.
   
\item[G.]
   Preserve in that license notice the full lists of Invariant Sections
   and required Cover Texts given in the Document's license notice.
   
\item[H.]
   Include an unaltered copy of this License.
   
\item[I.]
   Preserve the section Entitled ``History'', Preserve its Title, and add
   to it an item stating at least the title, year, new authors, and
   publisher of the Modified Version as given on the Title Page.  If
   there is no section Entitled ``History'' in the Document, create one
   stating the title, year, authors, and publisher of the Document as
   given on its Title Page, then add an item describing the Modified
   Version as stated in the previous sentence.
   
\item[J.]
   Preserve the network location, if any, given in the Document for
   public access to a Transparent copy of the Document, and likewise
   the network locations given in the Document for previous versions
   it was based on.  These may be placed in the ``History'' section.
   You may omit a network location for a work that was published at
   least four years before the Document itself, or if the original
   publisher of the version it refers to gives permission.
   
\item[K.]
   For any section Entitled ``Acknowledgements'' or ``Dedications'',
   Preserve the Title of the section, and preserve in the section all
   the substance and tone of each of the contributor acknowledgements
   and/or dedications given therein.
   
\item[L.]
   Preserve all the Invariant Sections of the Document,
   unaltered in their text and in their titles.  Section numbers
   or the equivalent are not considered part of the section titles.
   
\item[M.]
   Delete any section Entitled ``Endorsements''.  Such a section
   may not be included in the Modified Version.
   
\item[N.]
   Do not retitle any existing section to be Entitled ``Endorsements''
   or to conflict in title with any Invariant Section.
   
\item[O.]
   Preserve any Warranty Disclaimers.
\end{itemize}

If the Modified Version includes new front-matter sections or
appendices that qualify as Secondary Sections and contain no material
copied from the Document, you may at your option designate some or all
of these sections as invariant.  To do this, add their titles to the
list of Invariant Sections in the Modified Version's license notice.
These titles must be distinct from any other section titles.

You may add a section Entitled ``Endorsements'', provided it contains
nothing but endorsements of your Modified Version by various
parties---for example, statements of peer review or that the text has
been approved by an organization as the authoritative definition of a
standard.

You may add a passage of up to five words as a Front-Cover Text, and a
passage of up to 25 words as a Back-Cover Text, to the end of the list
of Cover Texts in the Modified Version.  Only one passage of
Front-Cover Text and one of Back-Cover Text may be added by (or
through arrangements made by) any one entity.  If the Document already
includes a cover text for the same cover, previously added by you or
by arrangement made by the same entity you are acting on behalf of,
you may not add another; but you may replace the old one, on explicit
permission from the previous publisher that added the old one.

The author(s) and publisher(s) of the Document do not by this License
give permission to use their names for publicity for or to assert or
imply endorsement of any Modified Version.

\begin{center}
{\Large\bf 5. COMBINING DOCUMENTS\par}
\phantomsection
\addcontentsline{toc}{section}{5. COMBINING DOCUMENTS}
\end{center}

You may combine the Document with other documents released under this
License, under the terms defined in section~4 above for modified
versions, provided that you include in the combination all of the
Invariant Sections of all of the original documents, unmodified, and
list them all as Invariant Sections of your combined work in its
license notice, and that you preserve all their Warranty Disclaimers.

The combined work need only contain one copy of this License, and
multiple identical Invariant Sections may be replaced with a single
copy.  If there are multiple Invariant Sections with the same name but
different contents, make the title of each such section unique by
adding at the end of it, in parentheses, the name of the original
author or publisher of that section if known, or else a unique number.
Make the same adjustment to the section titles in the list of
Invariant Sections in the license notice of the combined work.

In the combination, you must combine any sections Entitled ``History''
in the various original documents, forming one section Entitled
``History''; likewise combine any sections Entitled ``Acknowledgements'',
and any sections Entitled ``Dedications''.  You must delete all sections
Entitled ``Endorsements''.

\begin{center}
{\Large\bf 6. COLLECTIONS OF DOCUMENTS\par}
\phantomsection
\addcontentsline{toc}{section}{6. COLLECTIONS OF DOCUMENTS}
\end{center}

You may make a collection consisting of the Document and other documents
released under this License, and replace the individual copies of this
License in the various documents with a single copy that is included in
the collection, provided that you follow the rules of this License for
verbatim copying of each of the documents in all other respects.

You may extract a single document from such a collection, and distribute
it individually under this License, provided you insert a copy of this
License into the extracted document, and follow this License in all
other respects regarding verbatim copying of that document.

\begin{center}
{\Large\bf 7. AGGREGATION WITH INDEPENDENT WORKS\par}
\phantomsection
\addcontentsline{toc}{section}{7. AGGREGATION WITH INDEPENDENT WORKS}
\end{center}

A compilation of the Document or its derivatives with other separate
and independent documents or works, in or on a volume of a storage or
distribution medium, is called an ``aggregate'' if the copyright
resulting from the compilation is not used to limit the legal rights
of the compilation's users beyond what the individual works permit.
When the Document is included in an aggregate, this License does not
apply to the other works in the aggregate which are not themselves
derivative works of the Document.

If the Cover Text requirement of section~3 is applicable to these
copies of the Document, then if the Document is less than one half of
the entire aggregate, the Document's Cover Texts may be placed on
covers that bracket the Document within the aggregate, or the
electronic equivalent of covers if the Document is in electronic form.
Otherwise they must appear on printed covers that bracket the whole
aggregate.

\begin{center}
{\Large\bf 8. TRANSLATION\par}
\phantomsection
\addcontentsline{toc}{section}{8. TRANSLATION}
\end{center}

Translation is considered a kind of modification, so you may
distribute translations of the Document under the terms of section~4.
Replacing Invariant Sections with translations requires special
permission from their copyright holders, but you may include
translations of some or all Invariant Sections in addition to the
original versions of these Invariant Sections.  You may include a
translation of this License, and all the license notices in the
Document, and any Warranty Disclaimers, provided that you also include
the original English version of this License and the original versions
of those notices and disclaimers.  In case of a disagreement between
the translation and the original version of this License or a notice
or disclaimer, the original version will prevail.

If a section in the Document is Entitled ``Acknowledgements'',
``Dedications'', or ``History'', the requirement (section~4) to Preserve
its Title (section~1) will typically require changing the actual
title.

\begin{center}
{\Large\bf 9. TERMINATION\par}
\phantomsection
\addcontentsline{toc}{section}{9. TERMINATION}
\end{center}

You may not copy, modify, sublicense, or distribute the Document
except as expressly provided under this License.  Any attempt
otherwise to copy, modify, sublicense, or distribute it is void, and
will automatically terminate your rights under this License.

However, if you cease all violation of this License, then your license
from a particular copyright holder is reinstated (a) provisionally,
unless and until the copyright holder explicitly and finally
terminates your license, and (b) permanently, if the copyright holder
fails to notify you of the violation by some reasonable means prior to
60 days after the cessation.

Moreover, your license from a particular copyright holder is
reinstated permanently if the copyright holder notifies you of the
violation by some reasonable means, this is the first time you have
received notice of violation of this License (for any work) from that
copyright holder, and you cure the violation prior to 30 days after
your receipt of the notice.

Termination of your rights under this section does not terminate the
licenses of parties who have received copies or rights from you under
this License.  If your rights have been terminated and not permanently
reinstated, receipt of a copy of some or all of the same material does
not give you any rights to use it.

\begin{center}
{\Large\bf 10. FUTURE REVISIONS OF THIS LICENSE\par}
\phantomsection
\addcontentsline{toc}{section}{10. FUTURE REVISIONS OF THIS LICENSE}
\end{center}

The Free Software Foundation may publish new, revised versions
of the GNU Free Documentation License from time to time.  Such new
versions will be similar in spirit to the present version, but may
differ in detail to address new problems or concerns.  See
\texttt{http://www.gnu.org/copyleft/}.

Each version of the License is given a distinguishing version number.
If the Document specifies that a particular numbered version of this
License ``or any later version'' applies to it, you have the option of
following the terms and conditions either of that specified version or
of any later version that has been published (not as a draft) by the
Free Software Foundation.  If the Document does not specify a version
number of this License, you may choose any version ever published (not
as a draft) by the Free Software Foundation.  If the Document
specifies that a proxy can decide which future versions of this
License can be used, that proxy's public statement of acceptance of a
version permanently authorizes you to choose that version for the
Document.

\begin{center}
{\Large\bf 11. RELICENSING\par}
\phantomsection
\addcontentsline{toc}{section}{11. RELICENSING}
\end{center}

``Massive Multiauthor Collaboration Site'' (or ``MMC Site'') means any
World Wide Web server that publishes copyrightable works and also
provides prominent facilities for anybody to edit those works.  A
public wiki that anybody can edit is an example of such a server.  A
``Massive Multiauthor Collaboration'' (or ``MMC'') contained in the
site means any set of copyrightable works thus published on the MMC
site.

``CC-BY-SA'' means the Creative Commons Attribution-Share Alike 3.0
license published by Creative Commons Corporation, a not-for-profit
corporation with a principal place of business in San Francisco,
California, as well as future copyleft versions of that license
published by that same organization.

``Incorporate'' means to publish or republish a Document, in whole or
in part, as part of another Document.

An MMC is ``eligible for relicensing'' if it is licensed under this
License, and if all works that were first published under this License
somewhere other than this MMC, and subsequently incorporated in whole
or in part into the MMC, (1) had no cover texts or invariant sections,
and (2) were thus incorporated prior to November 1, 2008.

The operator of an MMC Site may republish an MMC contained in the site
under CC-BY-SA on the same site at any time before August 1, 2009,
provided the MMC is eligible for relicensing.

\backmatter
%%%%%%%%%%%%%%%%%%%%%%%%%%%%%%%%%%%%%%%%%%%%%%%%%%%%%%%%%%%%%%%%%
% Contents: The Bibliography
% File: biblio.tex (lshort2e.tex)
% $Id: biblio.tex 449 2010-12-14 16:53:51Z oetiker $
%%%%%%%%%%%%%%%%%%%%%%%%%%%%%%%%%%%%%%%%%%%%%%%%%%%%%%%%%%%%%%%%%
\bibliographystyle{plain}
\pagestyle{plain}
\bibliography{Cites}

%

% Local Variables:
% TeX-master: "lshort2e"
% mode: latex
% mode: flyspell
% End:

%\refstepcounter{chapter}
%\addcontentsline{toc}{chapter}{Index} 
%\printindex
\end{document}